\documentclass[sn-mathphys,Numbered]{sn-jnl}% Math and Physical Sciences Reference Style

\usepackage{graphicx}%
\usepackage{multirow}%
\usepackage{amsmath,amssymb,amsfonts}%
\usepackage{amsthm}%
\usepackage{mathrsfs}%
\usepackage[title]{appendix}%
\usepackage{xcolor}%
\usepackage{textcomp}%
\usepackage{manyfoot}%
\usepackage{booktabs}%
\usepackage{algorithm}%
\usepackage{algorithmicx}%
\usepackage{algpseudocode}%
\usepackage{listings}%

\begin{document}

\title[Article Title]{SPLICE  - Streamlining Digital Pathology \\ Image Processing}

\author{Areej Alsaafin} 
\author{Peyman Nejat} 
\author{Abubakr Shafique} 
\author{Jibran  Khan} 
\author{Saghir Alfasly} 
\author{Ghazal Alabtah} 
\author{H.R.Tizhoosh}

\affil{KIMIA Lab, Department of Artificial Intelligence \& Informatics, \\Mayo Clinic, Rochester, MN, USA}
%\affil[2]{Kimia Lab, University of Waterloo, Waterloo, ON, Canada}

\abstract{Digital pathology and the integration of artificial intelligence (AI) models have revolutionized histopathology, opening new opportunities. With the increasing availability of Whole Slide Images (WSIs), there's a growing demand for efficient retrieval, processing, and analysis of relevant images from vast biomedical archives. However, processing WSIs presents challenges due to their large size and content complexity. Full computer digestion of WSIs is impractical, and processing all patches individually is prohibitively expensive.

In this paper, we propose an unsupervised patching algorithm, Sequential Patching Lattice for Image Classification and Enquiry (SPLICE). This novel approach condenses a histopathology WSI into a compact set of representative patches, forming a ``collage" of WSI while minimizing redundancy. SPLICE prioritizes patch quality and uniqueness by sequentially analyzing a WSI and selecting non-redundant representative features. We evaluated SPLICE for search and match applications, demonstrating improved accuracy, reduced computation time, and storage requirements compared to existing state-of-the-art methods.

As an unsupervised method, SPLICE effectively reduces storage requirements for representing tissue images by 50\%. This reduction enables numerous algorithms in computational pathology to operate much more efficiently, paving the way for accelerated adoption of digital pathology.
} 
  
\maketitle

\section{Introduction}\label{introduction}
Modern computer vision algorithms have the potential to revolutionize the histopathology workflow. These algorithms possess the remarkable ability to process vast volumes of data, enabling pathologists to swiftly navigate and evaluate tissue slides. This not only expedites the diagnostic process but also minimizes errors in interpretation and classification. Deep networks inherently capture the essence of histopathology image content, encapsulating intricate patterns that a computer discerns from the image. They have the potential to uncover novel patterns and morphological features that might elude human observation~\cite{coudray2018classification,yu2016predicting}. These learned patterns, often referred to as \emph{feature representations}, are acquired during training and subsequently harnessed for diverse downstream applications. For instance, these feature representations can greatly facilitate the identification of similar histopathology images~\cite{kalra2020yottixel} through advanced image search and retrieval tools. This capability empowers pathologists to achieve enhanced diagnostic accuracy and reliability by harnessing computational pathology to explore extensive archives of previously diagnosed cases. However, most approaches developed for digital pathology require guidance from pathologists to obtain some representative samples from tumor regions that can be used to train a deep model. This is one of the main obstacles slowing down the development of machine learning in medical imaging as manual data annotation is a tedious and expensive process. Pathologists may spend hours annotating a few histopathology images, which in many cases are not enough to train a robust deep model. In addition, inter-observer variability is essentially one of the main issues in cancer diagnosis~\cite{zhang2019pathologist}. Therefore, multiple experts may be needed to annotate and analyze the same histopathology image to achieve more accurate labels for training deep networks.

WSIs represent an enormous computational challenge due to their gigapixel dimensions and complex patterns, making it infeasible for computers to process them in their entirety. The process of indexing whole slide images (WSIs) involves two primary tasks, each of which is of paramount importance: patching and deep feature extraction. In image processing, the patching stage involves dividing a WSI into patches, also referred to as ``tiles,'' which represent small-scale sub-images that are amenable to comprehensive computer analysis. The concept of ``Divide \& Conquer'' is a venerable strategy routinely used to address complex problems. It provides for the initial division of a complex problem into smaller, more manageable sub-problems, followed by conquering these component parts (in our case patches) in order to build back up a solution to solve the initial complex problem (in our case process a WSI). Due to the large size of WSIs, one has to split a WSI into more manageable patches to enable computerized processing, which is essentially the \emph{Divide} stage. However, this process has to be \emph{unsupervised} to be applicable to all types of tissue in all organs. As well, reliable patching should ensure comprehensive WSI coverage without inadvertently omitting any diagnostically relevant regions. This \emph{inclusivity} is essential to guarantee that the subsequent phase, the \emph{Conquer} stage, wherein the set of patches is fed into a deep network to represent the whole image. The accuracy of the diagnosis crucially depends on the quality of the WSI representation. All these have to be achieved by selecting as few patches as possible to reduce the computation time and storage requirements. 

\begin{figure}
    \centering
        \includegraphics[width=\textwidth,height=\textheight,keepaspectratio]{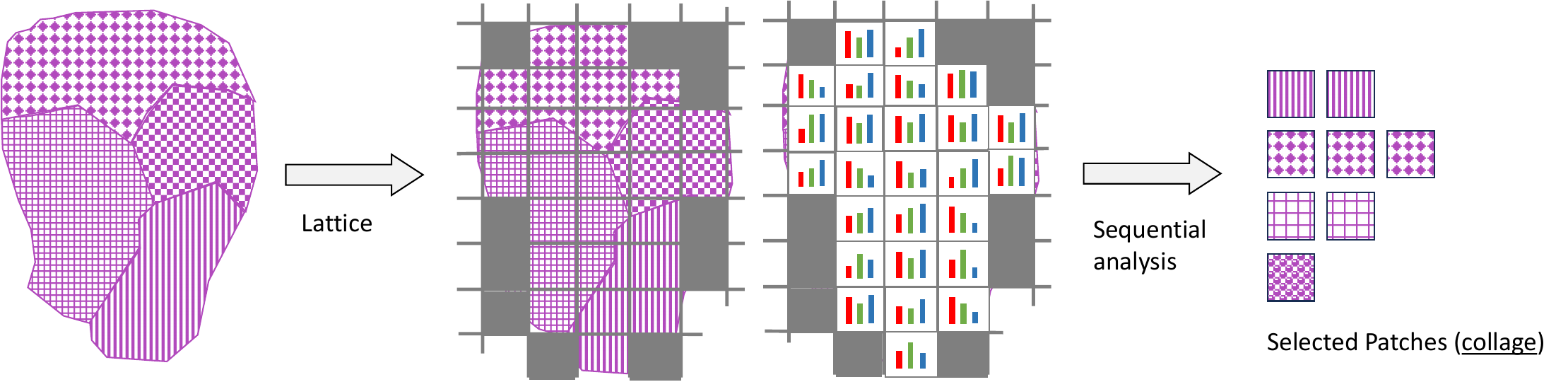}
        \caption{SPLICE approach for patch selection. SPLICE selects representative patches, a  ``collage'', by analyzing their color characteristics through sequential comparisons, including only patches with unique features in the WSI collage.}
        \label{fig:collage}
\end{figure}

This work introduces a novel patching algorithm called \emph{Sequential Patching Lattice for Image Classification and Enquiry} (SPLICE). At its core, SPLICE aims to sequentially analyze the patches of a WSI at low magnification and only keep one unique sample for each distinctive tissue morphology encountered (Figure \ref{fig:collage}). This strategy leads to forming a compact set of representative patches that we name a \emph{collage}. The design of the SPLICE is motivated by several crucial observations that collectively underscore the significance of its development. Firstly, there is a notable scarcity of solutions dedicated to representation learning and retrieval for entire high-resolution WSIs, with the majority of prior work focusing primarily on patch-level processing~\cite{hegde2019similar,schaer2019deep}. Additionally, the preponderance of research efforts has centered around labeled repositories, where malignancy regions are laboriously annotated by pathologists~\cite{iizuka2020deep,gecer2018detection,barker2016automated}. However, this fixation on labeled data can limit the potential of utilizing unlabelled, real-world data~\cite{wahab2022semantic}. Furthermore, the practicality of relying on real-valued features for image indexing has posed challenges due to their substantial storage and computational requirements~\cite{dubey2021decade,ai2013high}. Similarly, while hashing-based approaches can expedite search operations, they may not seamlessly facilitate data exchange among diverse repositories~\cite{jiang2016scalable,yang2017supervised}. Lastly, the inherent complexity of histopathology images, characterized by diverse edge shapes, intricate structures, stain variations, and high gradient changes, presents a formidable challenge for conventional computer vision algorithms~\cite{madabhushi2016image,komura2018machine}. Considering these observations, the SPLICE framework is designed to provide an unsupervised solution, addressing the intricacies of high-resolution WSIs while promoting efficiency and versatility in real-world applications.

\begin{figure}[htb]
    \centering
        \includegraphics[width=\textwidth,height=\textheight,keepaspectratio]{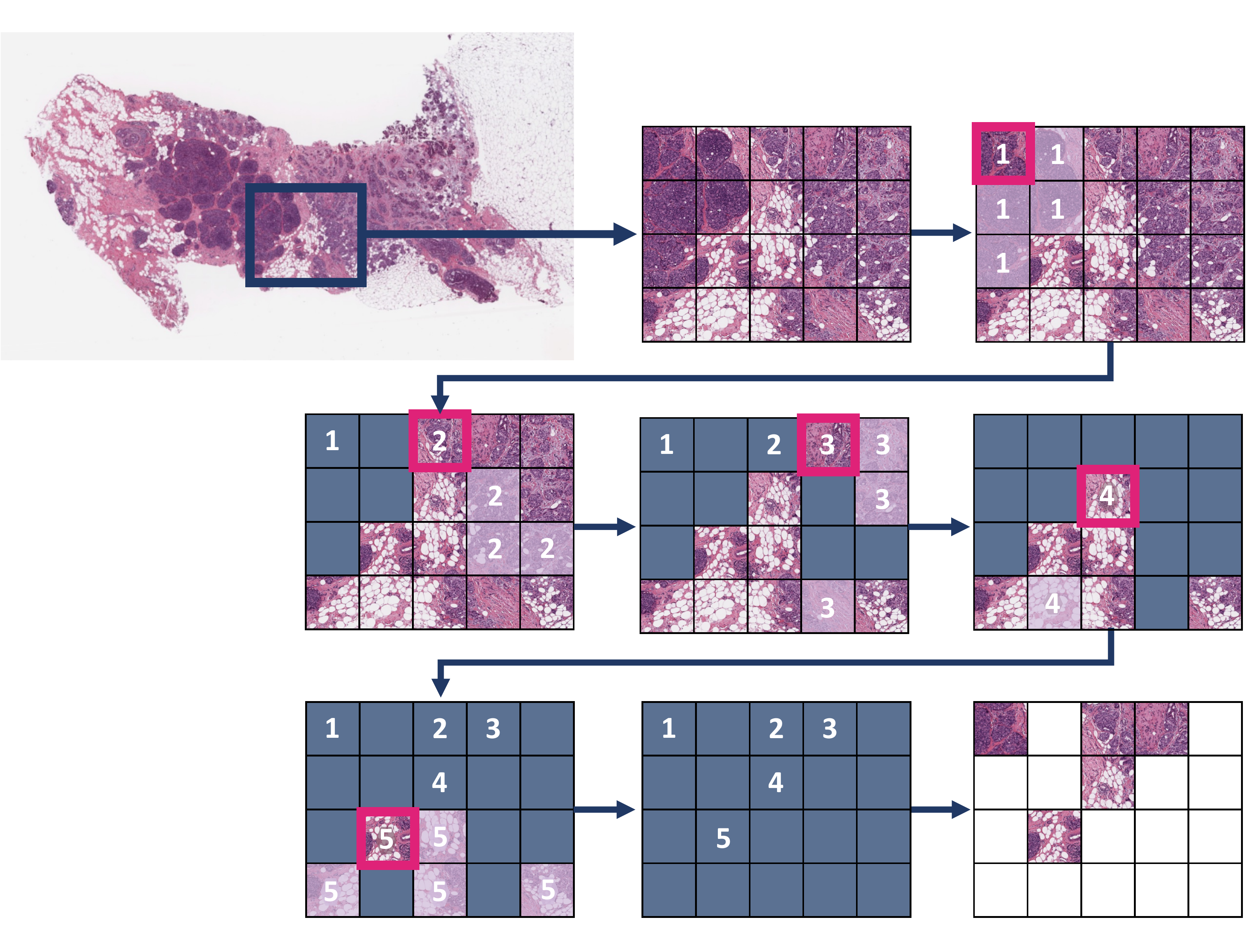}
        \caption{Simplified Example for SPLICE \emph{collage} applied on a WSI region (From top to bottom) -- Starting from an image, we define a lattice. In every pass, the reference patch (pointer) is marked with a pink square while the similar patches to the reference patch are overlaid. \textbf{First Pass:} Patch 1 selected and compared to all other patches, similar patches are marked by 1 and excluded, \textbf{Second Pass:} Patch 2 is compared to all remaining patches; similar patches are marked by 2 and excluded,  \textbf{Third Pass:} Patch 3 is compared to all remaining patches; similar patches are marked by 3 and excluded, \textbf{Fourth Pass:} Patch 4 is compared to all remaining patches; similar patches are marked by 4 and excluded, \textbf{Fifth Pass:} Patch 5 is compared to all remaining patches; similar patches are marked by 5 and excluded, \textbf{Last Pass:} the collage of selected patches is generated.}
        \label{fig:sequential}
\end{figure}

%----------------------------------------------------------------------
\section{Results}
%----------------------------------------------------------------------
The design of SPLICE (Figure \ref{fig:collage}) unveils two pivotal practical contributions. Firstly, an innovative approach is introduced, aiming to condense the entirety of a histopathology whole slide image into a unique set of patches, referred to as the \emph{collage}. This novel concept enables an efficient processing of gigapixel WSIs. Secondly, an end-to-end solution for the indexing and retrieval of WSIs based on their collage representation is proposed, providing a powerful and robust solution for image classification and retrieval with significant efficiency improvement compared to state-of-the-art patching solutions.

\begin{figure}[htb]
    \centering
        \includegraphics[width=\textwidth,height=\textheight,keepaspectratio]{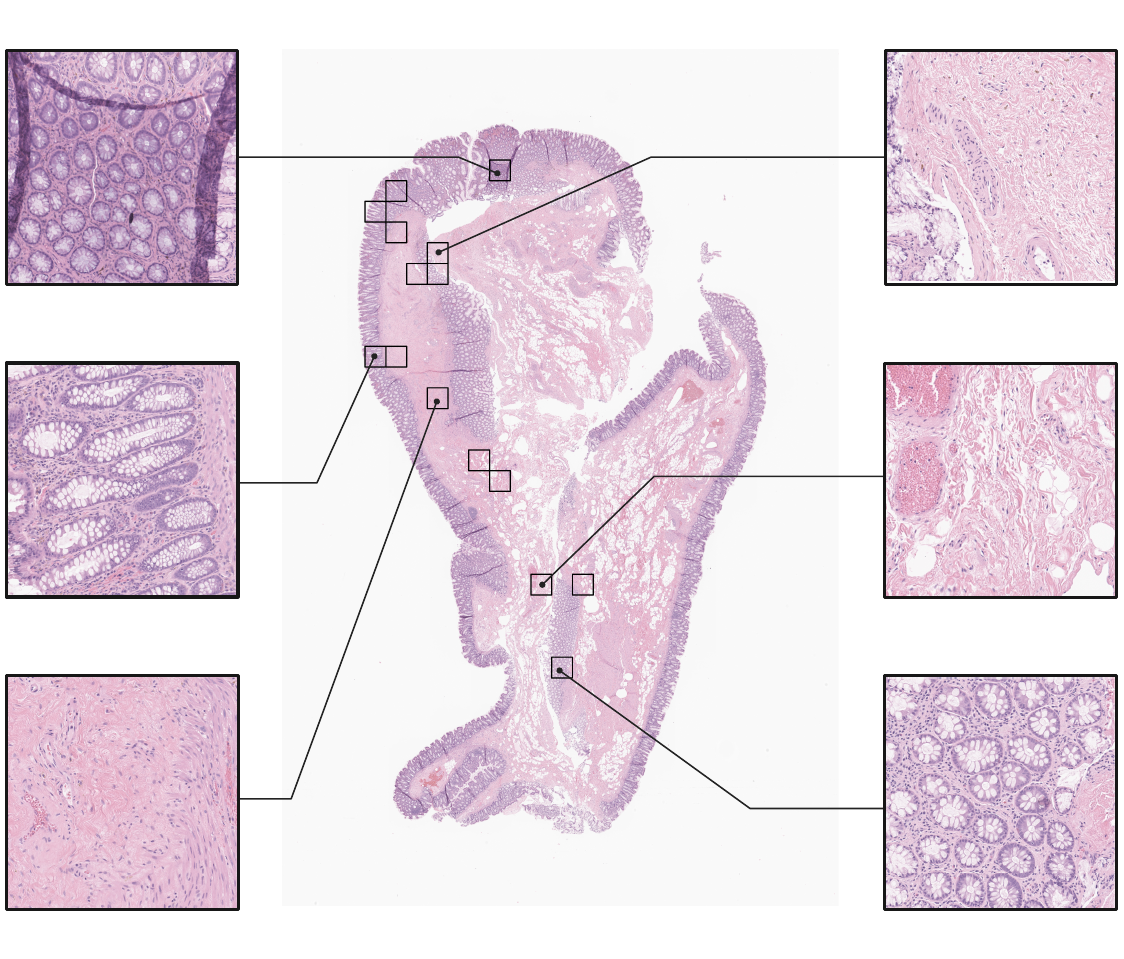}
        \caption{Representative patch selection using SPLICE in histopathology whole slide image from Mayo-CRC dataset. These diverse patches, extracted by SPLICE, showcase various patterns within the histopathology image of a colon specimen. The patches are selected by applying sequential analysis, leveraging the color characteristics of the patches for their distinctive representation.}
        \label{fig:samples}
\end{figure}

As shown in Figure~\ref{fig:sequential}, SPLICE is an approach that follows a systematic process to transform histopathology images into a compact collage representation, which preserves essential diagnostic information in a histopathology image while effectively mitigating both computational and analytical challenges. SPLICE can be used to process a queue of WSIs by calculating a compact representative collage for every WSI in the queue (a sample is shown in Figure~\ref{fig:samples}). The collage serves as a compact representation of the entire WSI that can be used to perform all downstream tasks, which in turn significantly alleviates the computation burden. 

\begin{figure}[htb]
    \centering
        \includegraphics[width=\textwidth,height=\textheight,keepaspectratio]{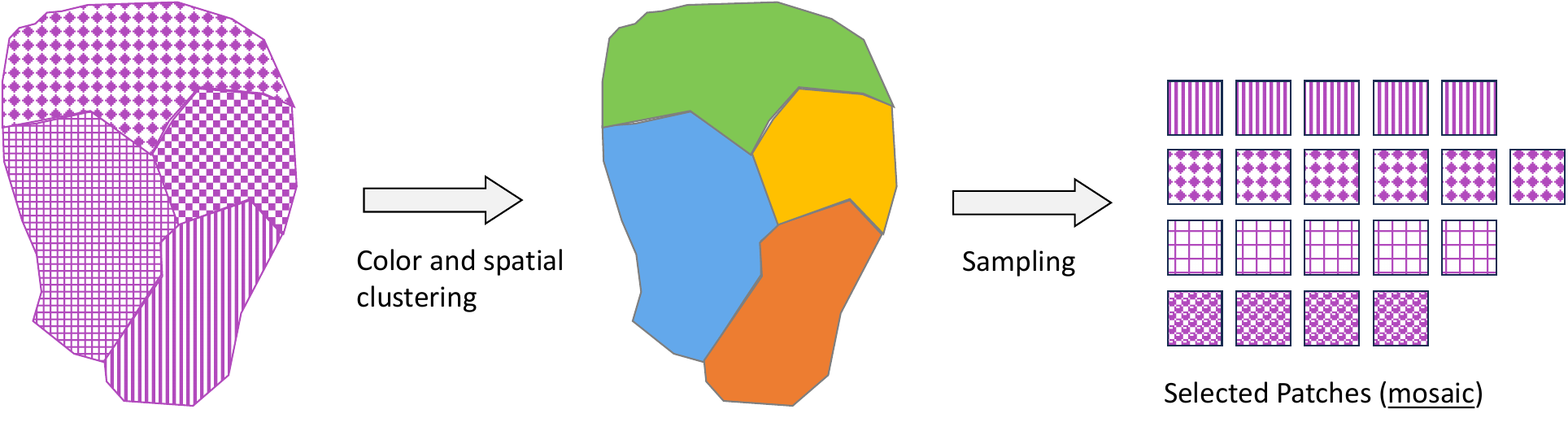}
        \caption{Yottixel approach for patch selection. Yottixel employs a two-step process, beginning with color clustering, followed by spatial clustering, to select a certain percentage of patches from each color cluster. The final set of selected patches is called the \emph{mosaic} that represents the WSI.}
        \label{fig:mosaic}
\end{figure}

The efficacy and efficiency of SPLICE are evaluated for image search and retrieval. This evaluation entails a comparison of search results achieved with SPLICE against those obtained using Yottixel search engine~\cite{kalra2020yottixel}, a state-of-the-art platform for patch selection and WSI matching. The primary objective of this evaluation is to facilitate rapid and efficient searching and retrieval of similar cases (both patches and WSIs) based on a given query image. As shown in Figure~\ref{fig:mosaic}, Yottixel employs a two-level k-means clustering approach, considering both color histogram and spatial proximity of patches. Consequently, a certain percentage of patches from each cluster are selected taking into account the intra-cluster heterogeneity. In contrast, SPLICE prioritizes patches with uniquely representative features. Yottixel represents a WSI image by generating a set of patches, referred to as a \emph{mosaic}. Whereas Yottixel's mosaic requires several parameters to be set (number of clusters, sample percentage), SPLICE works with only one parameter and does not need to make assumptions about the number of clusters. 

To assess the performance of SPLICE in search scenarios, various datasets were employed as detailed below. In each test scenario, SPLICE was employed for patching alongside Yottixel, yielding a \emph{mosaic} and a \emph{collage} for each WSI. In the case of the collage, the sequential analysis of WSI patches was performed utilizing their color histograms at a very low magnification level of 0.625x, using small patch dimensions of $32\times32$ pixels. This choice is based on empirical evidence indicating that higher magnifications do not significantly enhance performance for stain/color comparisons. Lower magnifications are favored as high resolutions primarily impact the shape and structure (morphological characteristics) in histopathology images. Subsequently, features are extracted from the patches belonging to both mosaic and collage using KimiaNet, a deep model with a DenseNet topology, trained on all diagnostic WSIs from TCGA (The Cancer Genome Atlas). The patch size employed in these experiments for feature extraction was set to $1024\times1024$ at 20x magnification as reported in literature \cite{kalra2020yottixel,kalra2020pan}. 

\begin{figure}
    \centering
        \includegraphics[width=\textwidth,height=\textheight,keepaspectratio]{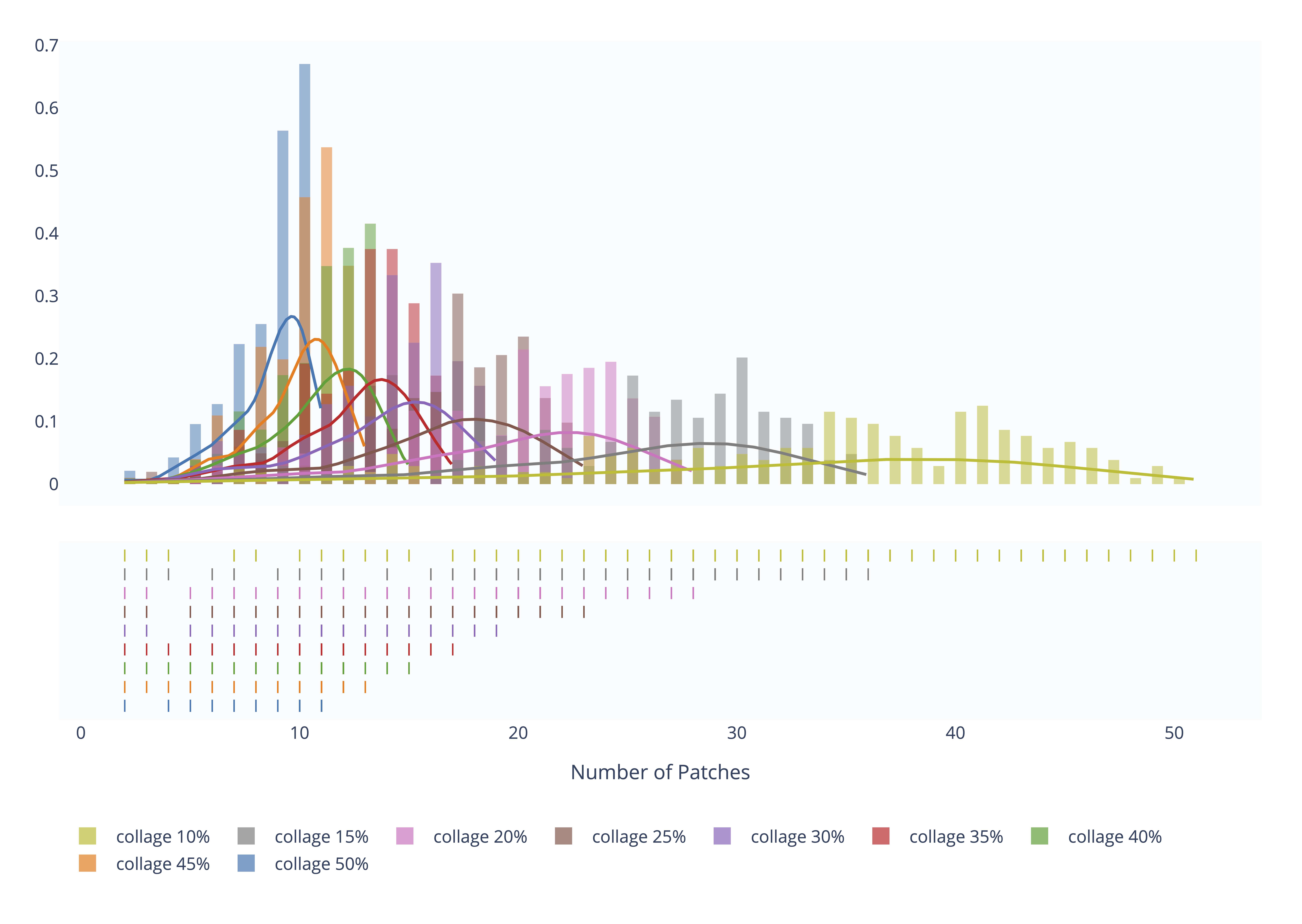}
        \caption{Distribution of patches across various \emph{collage} sizes, ranging from the 10th to the 50th percentile, employing whole-slide images sourced from the Mayo-CRC dataset. The histogram provides a visual representation of the patch distribution, while the rug plot at the bottom offers a detailed view of individual data points, enhancing the understanding of the data's spread and density.}
        \label{fig:collage_distribution}
\end{figure}

After generating the collage, comprising a low-redundant representative set of patches for each WSI, these patches were fed into KimiaNet to extract deep features. KimiaNet generates a feature vector of size 1024 for every patch in a given collage, resulting in $n$ feature vectors per WSI. For Yottixel, the mosaic was generated for each WSI using the configurations described in the original Yottixel paper~\cite{kalra2020yottixel}. These generated mosaic patches are subsequently processed in KimiaNet to extract features for each patch, yielding a set of mosaic feature vectors. After feature extraction, the resulting feature vectors were binarized using the \emph{MinMax} method~\cite{kumar2018deep, tizhoosh2015barcode}. The \emph{MinMax} algorithm maps each feature value to a binary string of 0s and 1s by applying an approximation of 1D derivatives of feature values. This approach enables fast Hamming distance search by comparing the binary barcodes, instead of the real values of feature vectors through Euclidean distance. A set of binary barcodes is obtained for each WSI to represent its mosaic/collage feature vectors. Hence, each WSI in the archive is represented by a set of binary barcodes. 

To perform the evaluation, the leave-one-out validation was employed to compare the set of binary barcodes of every image (as a query image) to those of the images stored in the archive. The Hamming distance between the barcodes of the query image and the barcodes of every image in the archive was calculated. To compare sets of patches, the ``median of minimum'' approach, introduced in Yottixel~\cite{kalra2020yottixel}, was performed to enable WSI-to-WSI matching. This method computes the distance between two WSIs based on the median of the minimum Hamming distances among the barcodes of their patch sets. In essence, the distance between a query WSI ($I_q$) and another WSI ($I$) is determined by the median of the minimum Hamming distances between each barcode in the patch set of $I_q$ and the patch set of $I$. The best match is identified as the one with the median value of the minimum Hamming distances among the patches (i.e., the barcodes) of the two sets. Based on the calculated distances, the top-1 match and the \emph{majority vote} of top-3 (MV@3) and top-5 (MV@5) matches were retrieved for each WSI. Majority vote criterion indicates that at least $n/2 + 1$ of the top-$n$ images should belong to the same class as the query image. Majority voting is much more reliable than top-n accuracy common in computer vision \cite{kalra2020pan}

The following sections detail the experiments conducted on three different datasets to evaluate the performance of SPLICE compared to Yottixel's mosaic. Two of the datasets are private datasets obtained from Mayo Clinic, which are Mayo-CRC and Mayo-Liver. The third dataset is the public TCGA with different primary sites.

\begin{figure}[htb]
    \centering
        \includegraphics[width=\textwidth,height=\textheight,keepaspectratio]{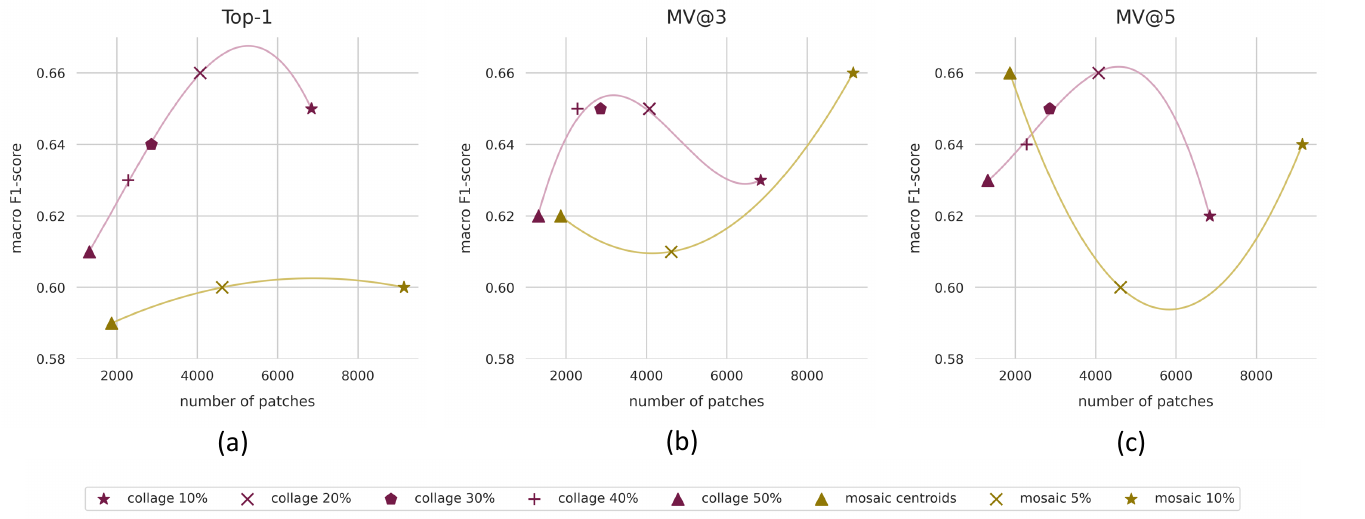}
        \caption{Macro F1-scores for top-k image retrieval in the Mayo-CRC dataset, showcasing the impact of varying \emph{collage} and \emph{mosaic} sizes. The diagrams illustrate the outcomes derived from leave-one-out and majority vote (MV) techniques for three specific scenarios: (a) Top-1, (b) MV@3, and (c) MV@5. The \emph{mosaic} thresholds employed include 9 patches, or 9 centroids (the smallest possible number of patches by mosaic), 5\%, and 10\%, while the \emph{collage} thresholds encompass the 10th, 20th, 30th, 40th, and 50th percentiles.}
        \label{fig:collage_mosaic_analysis}
\end{figure}

\subsection{WSI Search \& Match Using Mayo Clinic Datasets}
Two private datasets were obtained from Mayo Clinic and were utilized to evaluate the performance of SPLICE in image search applications. The first dataset, Mayo-CRC, comprises a collection of histopathology images from colorectal cancer cases treated at Mayo Clinic. In the Mayo-CRC dataset, 63 WSIs were labeled as Cancer-Adjacent Polyps (CAP), another 63 were labeled as Non-Recurrent Polyps (POP-NR), and 83 were labeled as Recurrent Polyps (POP-R). Note that labels are at WSI level not at pixel level. The second dataset, referred to as Mayo-Liver, encompasses histopathology images from fatty liver disease cases at Mayo Clinic. Mayo-Liver includes three classes: 150 WSIs belonging to Alcoholic Steatohepatitis (ASH), 158 WSIs attributed to Non-Alcoholic Steatohepatitis (NASH), and 18 WSIs categorized as Normal. For both the Mayo-CRC and Mayo-Liver datasets, high-throughput slide scans were conducted at a magnification of 40x within the facilities of Mayo Clinic Rochester Pathology Research Core, utilizing scanning equipment such as the Aperio AT Turbo Scanner, Aperio AT2 Scanner, and Aperio GT450 Scanner. All slide scans are for slides that contain Formalin-Fixed Paraffin-Embedded (FFPE) tissue specimens. 

Yottixel's mosaic and SPLICE take distinct approaches to the patch selection process. Yottixel's mosaic prioritizes well-distributed patch selection from all regions of a WSI, whereas SPLICE places a higher emphasis on the uniqueness of the selected patches. Yottixel's mosaic original configuration, as detailed in~\cite{kalra2020yottixel}, employs a k value of 9 for color k-means clustering, while the spatial k-means clustering is set at 0.05, which corresponds to 5\% of patches in every color cluster. Increasing these values results in the selection of a larger number of patches per WSI. When it comes to the collage generation, the threshold is dynamically determined based on the population of patches within each WSI. This threshold adapts continuously during each iteration of the sequential scanning process, depending on the remaining patches in the pool. Therefore, the threshold is expressed as a percentile of the calculated distances in each iteration, rather than as a fixed numerical value. Figure~\ref{fig:collage_distribution} provides insight into the distribution of the number of patches per WSI across different percentiles, ranging from the 10th to the 50th percentile (the median). A higher percentile corresponds to a more stringent threshold, resulting in fewer patches included in the collage of each WSI. Conversely, a lower percentile implies a more lenient threshold, accommodating more patches. However, as demonstrated in Figure~\ref{fig:collage_distribution}, using a strict threshold, such as the 10th percentile, may lead to a small number of patches for some WSIs, either due to a limited number of patches (e.g., for small tissue fragments) or because of the high similarity between the patches, resulting in a reduced number of representative patches. A visual representation of the selected patches is shown in Figure~\ref{fig:samples}. The figure shows the selected patches for a WSI with a colon specimen from Mayo-CRC. The patches were selected by applying sequential analysis with the 25th percentile as a similarity threshold for the color characteristics of the patches.

%-------------------------
\begin{figure}[htb]
    \centering
        \includegraphics[width=\textwidth,height=\textheight,keepaspectratio]{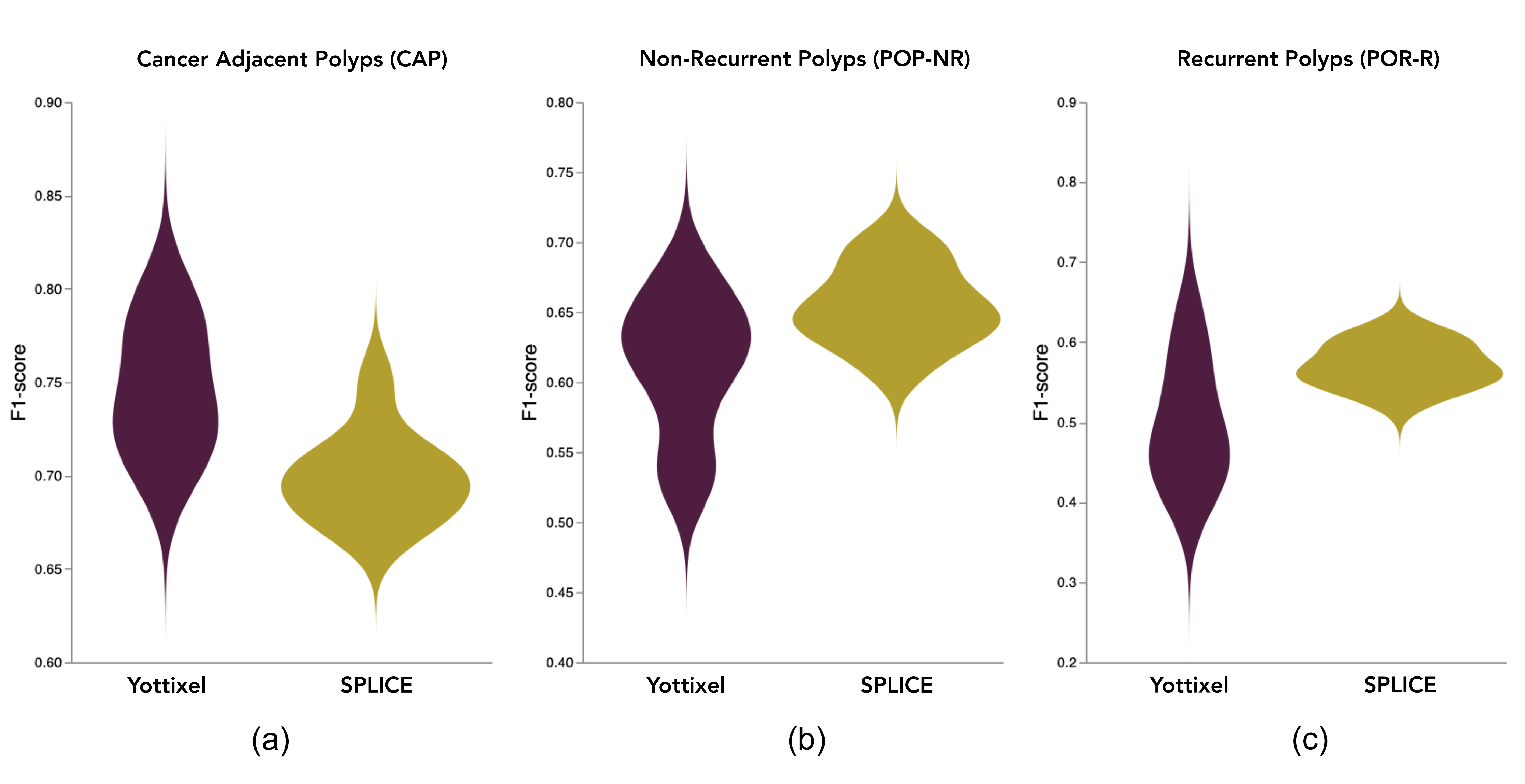}
        \caption{Distribution of macro F1-scores, comparing SPLICE and Yottixel's mosaic, across various majority-vote scenarios, including top-1, top-3, and top-5, for each class within the Mayo-CRC dataset. These classes consist of (a) Cancer-Adjacent Polyps, (b) Non-Recurrent Polyps, and (c) Recurrent Polyps.}
        \label{fig:violin}
\end{figure}
% %-------------------------

An extensive experiment was conducted to investigate the impact of varying percentiles for SPLICE in comparison to adjusting the spatial parameter of Yottixel's mosaic. In this experimental analysis, for SPLICE, five thresholds, representing the 10th, 20th, 30th, 40th, and 50th percentiles, were explored. The three thresholds considered for Yottixel's mosaic encompass 9 patches (color cluster centroids), 5\% (following the original work's configuration), and 10\% to facilitate the inclusion of a higher number of patches per WSI. 

Figure~\ref{fig:collage_mosaic_analysis} presents the macro F1-scores for retrieving the top-1, majority vote top-3 (MV@3), and majority vote top-5 (MV@5) similar WSIs for each WSI in the Mayo-CRC dataset, employing leave-one-out evaluation. The figure also shows the number of patches selected for generating the \emph{collage} and \emph{mosaic} for each WSI within the dataset. The highest macro F1-score achieved by SPLICE for top-1 was 0.66 when using the 20th percentile for \emph{collage} creation, with 4,073 patches extracted from 209 WSIs in the Mayo-CRC dataset. In comparison, Yottixel's mosaic attained a macro F1-score of 0.60 for top-1 when employing 5\% and 10\% thresholds, with a total number of patches amounting to 4,619 and 9,143, respectively. Remarkably, both SPLICE and Yottixel's mosaic exhibited similar performance with a macro F1-score of 0.66 for MV@3 under the most strict thresholds in this experiment (generating the fewest patches for both methods), with 1,814 and 1,868 total patches for SPLICE and Yottixel, respectively. In all measures shown in Figure~\ref{fig:collage_mosaic_analysis}, the \emph{collage} with the 20th percentile and the \emph{mosaic} with 5\% thresholds result in a comparable number of patches, but SPLICE displayed better performance across top-1, MV@3, and MV@5 by 4\% to 6\%. It's essential to consider that when employing MV@5, mosaic utilized more than four times as many patches (9,143) as SPLICE (2,071) to achieve the same level of performance. Overall, SPLICE demonstrated better performance across different percentiles compared to the various thresholds used for mosaic. The general trend for Yottixel across different thresholds is that more patches tend to improve performance, especially for top-1 and MV@3. Additionally, using more patches to generate the \emph{mosaic} had a smaller impact on the retrieval results of top-1 compared to MV@3 and MV@5, indicating that more patches per WSI were needed to achieve a correct vote for the retrieved similar WSIs. On the other hand, including more patches to generate the \emph{collage} resulted in better results in identifying the most similar WSI (top-1), suggesting that SPLICE selects representative and less redundant patches. The performance for MV@3 and MV@5 also improved with more patches until the 20th percentile, when the macro F1-score began to decrease as the criteria to include or exclude patches from a \emph{collage} became less strict. In general, there is a trade-off between performance on one side, and computation and storage requirements on the other side. However, SPLICE mitigated the impact of this trade-off by achieving satisfactory performance with strict thresholds that allow a small number of patches to be used in generating WSI \emph{collages}.

The violin plots in Figure~\ref{fig:violin} depict the distribution of F1-scores for the top-1, top-3, and top-5 majority vote scenarios across three distinct thresholds applied to each class within the Mayo-CRC dataset. For mosaic, 9-patches, 5\%, and 10\% were considered, while for SPLICE, 20th, 30th, and 4th percentiles were included in this experiment. A noticeable pattern emerges as the distribution of F1-scores for Yottixel exhibits a more pronounced vertical spread, reflecting sensitivity to varying threshold values. In contrast, SPLICE's distribution is characterized by a more horizontal orientation, indicating a more consistent performance, with the threshold having a less pronounced impact. This difference underscores the quality of images incorporated in the \emph{collage} generated by SPLICE, which contributes to stable and reliable performance across a range of threshold values. The Yottixel's mosaic showed better performance in identifying images from the CAP class when a higher number of patches were included. In contrast, SPLICE demonstrated constant performance in categorizing images from the POP-NR and POP-R classes.

\begin{table}
\centering
\caption{Comparison of the performance of SPLICE's \emph{collage}, and Yottixel's \emph{mosaic} using CRC dataset from Mayo Clinic. The evaluation is conducted using majority vote and leave-one-out validation. The table displays the results for majority vote of \textit{top-1}, \textit{top-3}, and \textit{top-5} in terms of accuracy, precision, recall, and macro F1-score metrics.}
\label{tab:crc_search}
    \begin{tabular*}{\textwidth}{l @{\extracolsep{\fill}} lccccc}
    \toprule
    \textbf{Method} & \textbf{Metric} & \textbf{Top-1} & \textbf{MV@3} & \textbf{MV@5}  \\
    \midrule
    \multirow{4}{*}{SPLICE}     & Accuracy    & 0.62 & 0.65 & 0.63   \\
                                & Precision   & 0.63 & 0.65 & 0.64   \\
                                & Recall      & 0.63 & 0.66 & 0.64   \\
                                & F1-score    & 0.63 & 0.65 & 0.64   \\
    \midrule
    \multirow{4}{*}{Yottixel's mosaic}   & Accuracy    & 0.60  & 0.60 & 0.60  \\
                                & Precision   & 0.62 & 0.61 & 0.61   \\
                                & Recall      & 0.62 & 0.62 & 0.63   \\
                                & F1-score    & 0.60 & 0.61 & 0.60   \\
    \bottomrule
    \end{tabular*}
\end{table}

Table~\ref{tab:crc_search} provides a comprehensive performance comparison between collage and mosaic, with a focus on various metrics spanning top-1, top-3, and top-5 majority vote scenarios, assessed using leave-one-out evaluation on the Mayo-CRC dataset. These results were generated by employing the 40th percentile threshold for SPLICE and a 5\% threshold for Yottixel's mosaic. The findings depicted in Table~\ref{tab:crc_search} demonstrate that SPLICE outperforms Yottixel's mosaic across all metrics for each of the three criteria: top-1, top-3, and top-5. For instance, in the context of top-3 majority vote, SPLICE attains an accuracy of 65\%, surpassing mosaic's search accuracy of 60\%. Furthermore, SPLICE exhibits recall of 66\% and precision of 65\% in comparison to mosaic's recall of 62\% and precision of 61\% for the same top-3 criterion.

\begin{table}
\centering
\caption{Comparison of the performance of SPLICE's \emph{collage}, and Yottixel's \emph{mosaic} using Liver dataset from Mayo Clinic. The evaluation is conducted using majority vote and leave-one-out validation. The table displays the results for the majority vote of \textit{top-1}, \textit{top-3}, and \textit{top-5} in terms of accuracy, precision, recall, and macro F1-score metrics.}
\label{tab:liver_search}
    \begin{tabular*}{\textwidth}{l @{\extracolsep{\fill}} lccccc}
    \toprule
    \textbf{Method} & \textbf{Metric} & \textbf{Top-1} & \textbf{MV@3} & \textbf{MV@5}  \\
    \midrule
    \multirow{4}{*}{SPLICE}     & Accuracy    & 0.83 & 0.79 & 0.77   \\
                                & Precision   & 0.83 & 0.87 & 0.85   \\
                                & Recall      & 0.74 & 0.63 & 0.66   \\
                                & F1-score    & 0.77 & 0.68 & 0.71   \\

    \midrule
    \multirow{4}{*}{Yottixel's mosaic}   & Accuracy    & 0.76 & 0.79 & 0.80  \\
                                & Precision   & 0.76 & 0.87 &  0.88 \\
                                & Recall      & 0.59 &  0.63 & 0.62 \\
                                & F1-score    & 0.62 & 0.67 & 0.65  \\
    \bottomrule
    \end{tabular*}
\end{table}

The assessment was extended to the Mayo-Liver dataset, with results presented in Table~\ref{tab:liver_search}, again employing leave-one-out evaluation across top-1, top-3, and top-5 majority vote scenarios. These results were generated by employing the 35th percentile threshold for SPLICE and a 5\% threshold for Yottixel's mosaic. SPLICE showcases a markedly better macro F1-score across all criteria, exhibiting a substantial 15\% difference for top-1, achieving an impressive 77\% in contrast to mosaic's 62\%. In terms of accuracy for top-1, SPLICE outperforms Yottixel's mosaic by 7\%, achieving an accuracy rate of 83\% as compared to mosaic's 76\%. However, the mosaic demonstrates higher accuracy when the top-5 criterion is applied, reaching 80\%, surpassing SPLICE's 77\%. However, it's interesting to observe that collage and mosaic exhibit remarkably similar performance when assessed under the top-3 majority vote scenario. Of particular interest is the observation that Ymosaic's performance seems to improve as the number of retrieved images increases, while SPLICE excels at retrieving the most similar top-1 image, demonstrating robust performance across accuracy, recall, precision, and macro F1-score metrics.

\begin{figure}
    \centering
        \includegraphics[width=\textwidth,height=\textheight,keepaspectratio]{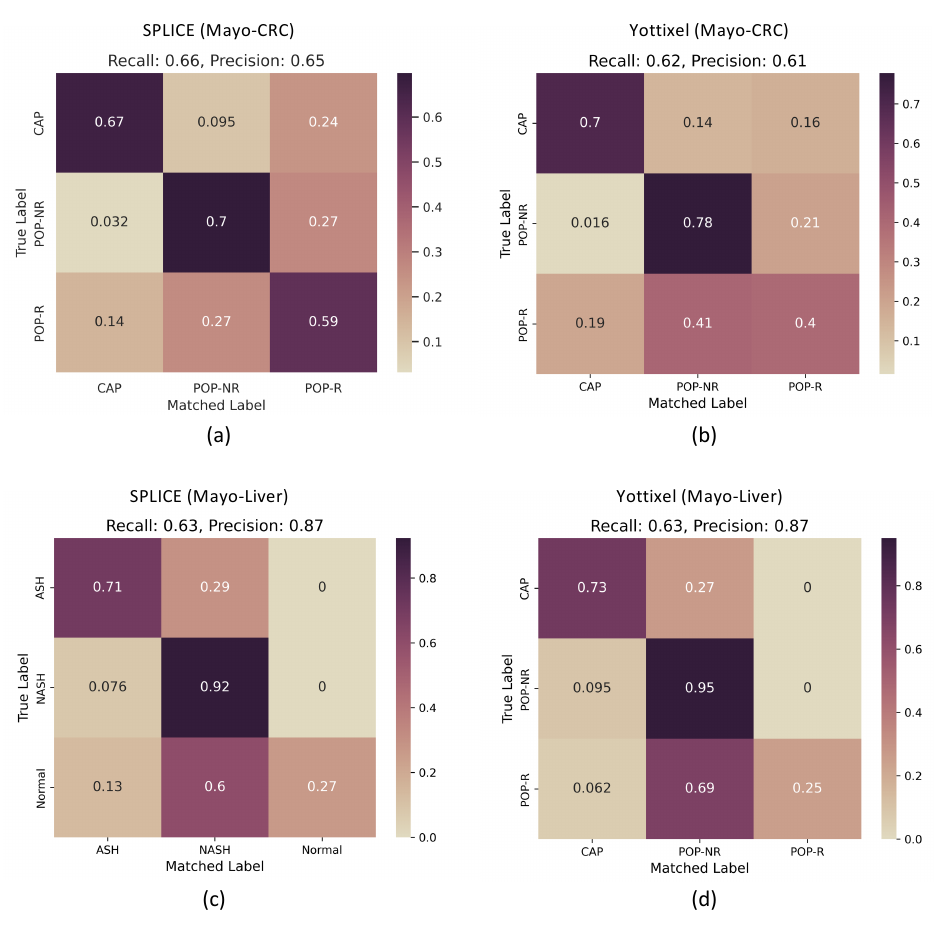}
        \caption{Confusion matrices for search and match in the Mayo-CRC dataset with SPLICE's \emph{collage} representation (a), Yottixel's \emph{mosaic} representation with Mayo-CRC (b), SPLICE on Mayo-Liver (c), and Yottixel's mosaic with Mayo-Liver (d). These matrices depict the search results obtained through leave-one-out validation and top-3 majority vote techniques for colorectal cancer and fatty liver disease datasets.}
        \label{fig:cm_crc_liver}
\end{figure}

Figure~\ref{fig:cm_crc_liver} presents the confusion matrices of the search results achieved by SPLICE's \emph{collage} and Yottixel's \emph{mosaic} representations, specifically when applied to the Mayo-CRC and Mayo-Liver datasets. The matrices were generated through leave-one-out validation. Additionally, the matrices illustrate the outcomes of the top-3 majority vote for each image in both colorectal and fatty liver disease datasets.

\subsection{WSI Search \& Match Using Public Datasets}
\begin{figure}
    \centering
        \includegraphics[width=\textwidth,height=\textheight,keepaspectratio]{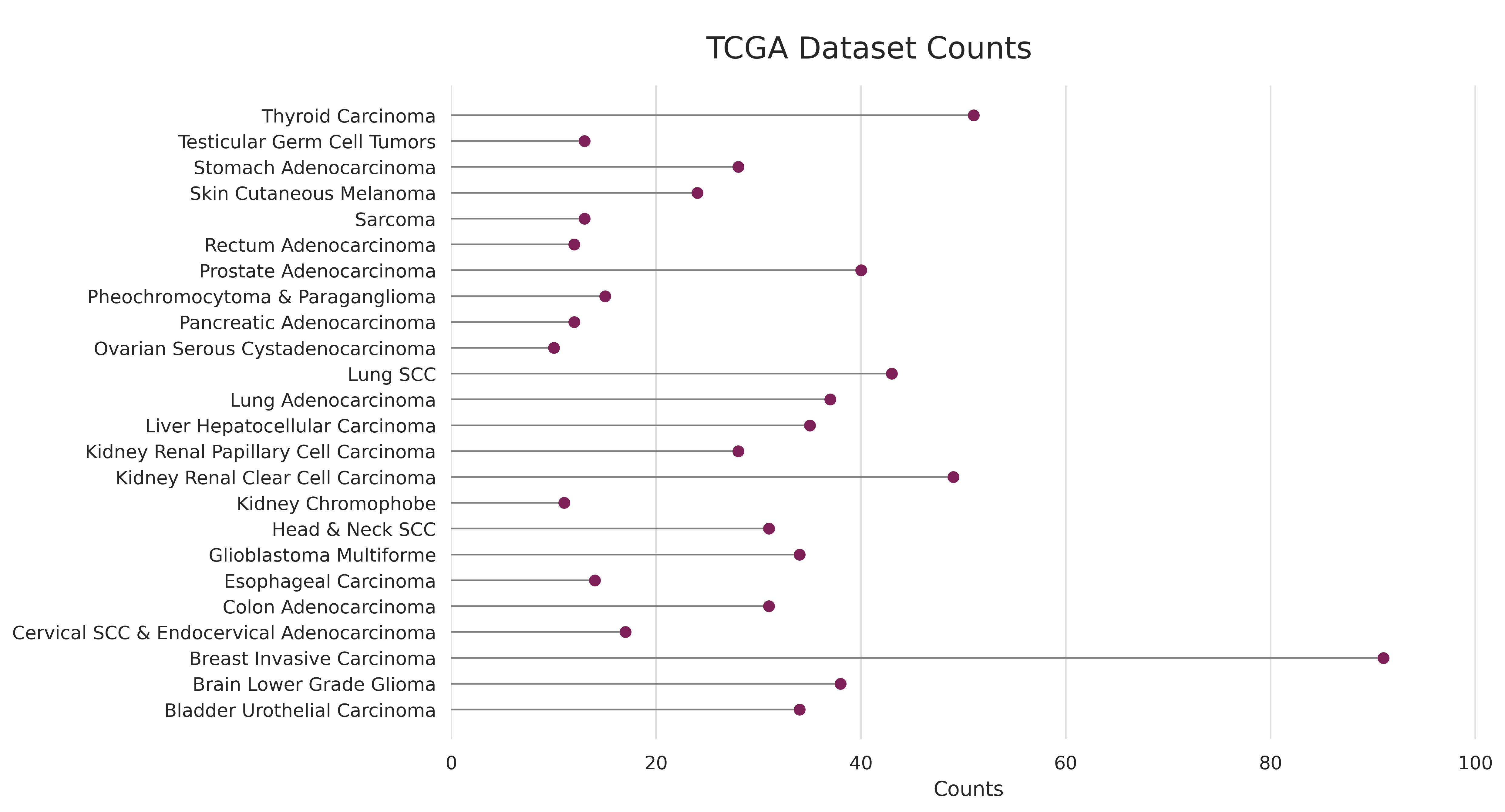}
        \caption{Visual representation of TCGA dataset counts for diverse cancer types.}
        \label{fig:tcga_counts}
\end{figure}

SPLICE's performance was further assessed using public datasets obtained from The Cancer Genome Atlas (TCGA) project, which is publicly available through the National Cancer Institute (NCI). As depicted in Figure~\ref{fig:tcga_counts}, this assessment included a total of 711 WSIs derived from 24 TCGA datasets originating from 20 primary sites. These WSIs contained Formalin-Fixed Paraffin-Embedded (FFPE) tissue specimens. These WSIs from TCGA have been utilized for KimiaNet's testing. Therefore, to ensure an unbiased assessment, all cases that were part of the training set of KimiaNet were omitted from the evaluation.

Table~\ref{tab:tcga_search} presents a comparative analysis of the performance between SPLICE's \emph{collage} and Yottixel's \emph{mosaic} using the TCGA dataset. In this experiment, a 5\% threshold was applied for Yottixel's mosaic, while the 30th percentile was employed for SPLICE. The evaluation was carried out using majority vote and leave-one-out validation. Table~\ref{tab:tcga_search} provides a comprehensive overview of the results for top-1, MV@3, and MV@5 in terms of accuracy, precision, recall, and macro F1-score metrics. The results in Table~\ref{tab:tcga_search} demonstrate that collage and mosaic generated comparable results, utilizing 11,663 patches for SPLICE and 25,679 patches for Yottixel's mosaic. Across most evaluation metrics, SPLICE outperforms Yottixel's mosaic, displaying higher accuracy, precision, recall, and macro F1-scores. However, for MV@3, both methods achieve an identical accuracy of 0.78. This comparison underscores the effectiveness of SPLICE, even when using significantly fewer patches than Yottixel's mosaic. In fact, less than half of the patches used by Yottixel's mosaic were used by SPLICE to generate \emph{collage} representations for WSIs. 

\begin{figure}
    \centering
        \includegraphics[width=\textwidth,height=\textheight,keepaspectratio]{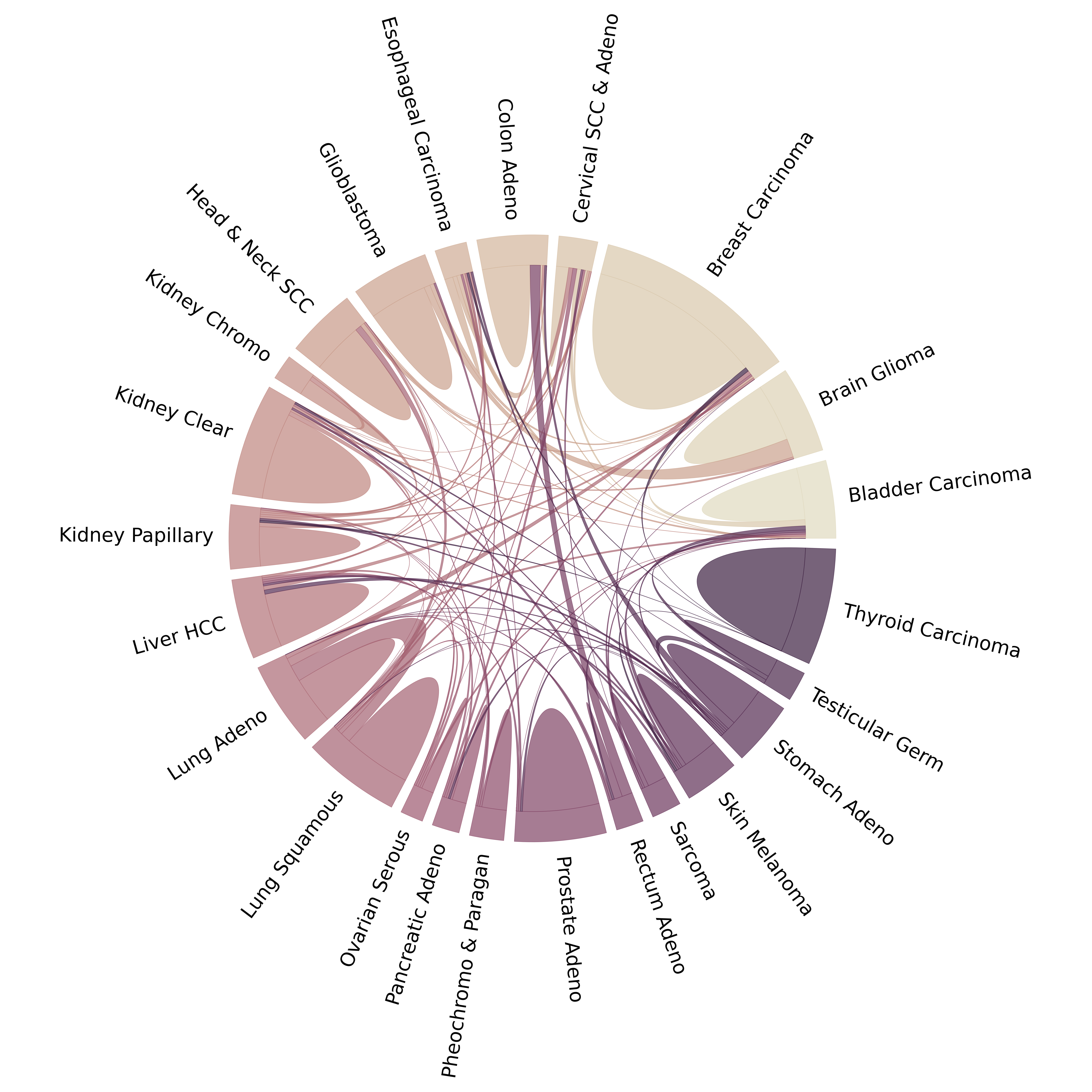}
        \caption{Chord diagram illustrating the TCGA dataset search results using leave-one-out and majority vote of top 3 matches with \emph{collage} representations generated using SPLICE. The diagram shows the interplay of different cancer classes within the TCGA dataset. Each chord connection depicts the relationships between cancer classes when employing search and match with a majority vote of the top 3 similar whole slide images.}
        \label{fig:tcga_chord}
\end{figure}

Figure~\ref{fig:tcga_chord} presents a chord diagram illustrating the TCGA dataset search results using leave-one-out and majority vote of top 3 matches with \emph{collage} representations generated using SPLICE. The diagram shows the interplay of different cancer classes within the TCGA dataset. Each chord connection depicts the relationships between cancer classes when employing search and match with a majority vote of the top 3 similar whole slide images. This visualization provides insight into the performance of SPLICE in categorizing and retrieving WSIs from a diverse set of cancer classes within the TCGA dataset.

%-------------------------
%-------------------------

\begin{table}
\centering
\caption{Comparison of the performance of SPLICE's \emph{collage}, and Yottixel's \emph{mosaic} using TCGA dataset. The evaluation is conducted using majority vote and leave-one-out validation. The table displays the results for the majority vote of \textit{top-1}, \textit{top-3}, and \textit{top-5} in terms of accuracy, precision, recall, and macro F1-score metrics.}
\label{tab:tcga_search}
    \begin{tabular*}{\textwidth}{l @{\extracolsep{\fill}} lccccc}
    \toprule
    \textbf{Method} & \textbf{Metric} & \textbf{Top-1} & \textbf{MV@3} & \textbf{MV@5}   \\
    \midrule
    \multirow{4}{*}{SPLICE} & Accuracy & 0.77 & 0.78 & 0.78  \\
                          & Precision & 0.76 & 0.79  & 0.79  \\
                          & Recall & 0.72 & 0.73 &  0.71 \\
                          & F1-score & 0.74 & 0.74 & 0.73  \\
    \midrule
    \multirow{4}{*}{Yottixel's mosaic} & Accuracy & 0.76 & 0.78 & 0.77  \\
                              & Precision & 0.73 & 0.75 &  0.74 \\
                              & Recall & 0.71 & 0.71 &  0.69 \\
                              & F1-score & 0.71 & 0.73 & 0.71  \\
    \bottomrule
    \end{tabular*}
\end{table}

%-------------------------
%-------------------------

The primary objective of SPLICE is to select representative patches from WSIs, exhibiting low redundancy, for \emph{collage} generation, striking a balance between a minimal number of patches and high performance. In contrast, Yottixel excels in retrieving similar WSIs through the generation of \emph{mosaics}, which inherently involve a higher number of patches even when a small threshold (like 5\%) is used. This distinction is prominently evident in Table~\ref{tab:search_time}, which provides insights into the total number of patches required to generate \emph{collages} and \emph{mosaics} across the three datasets considered in the experiments: Mayo-CRC, Mayo-Liver, and TCGA. It's important to note that the configurations for both Yottixel's mosaic and SPLICE to obtain these results align with those used in Table~\ref{tab:crc_search}, Table~\ref{tab:liver_search}, and Table~\ref{tab:tcga_search}.

\begin{table}[htb]
\centering
\caption{Comparison of the number of selected patches for SPLICE's \emph{collage} and Yottixel's \emph{mosaic}. The search time varies for each method depending on the number of patches used to represent each WSI. A higher number of selected patches may lead to longer processing time for searching and retrieving similar cases to a query image. The storage requirement also increases for a larger number of patche embeddings.}
\label{tab:search_time}
\begin{tabular}{lllccccc}
\toprule
& \textbf{patching} & \textbf{\# patches} & \textbf{\# patches/WSI} & \textbf{storage (KB)} & \textbf{time (sec)}  \\
\midrule

    \multirow{2}{*}{Mayo-CRC}     & collage     & 2,282     & 10 ± 2     & 43 ± 10     &   6.01      \\ % 0.02 ± 0.02
                                 & mosaic      & 4,619     & 22 ± 15    & 96 ± 51    &   16.89     \\  % 0.04 ± 0.06
    \bottomrule
    
    \multirow{2}{*}{Mayo-Liver}   & collage     & 2,709     & 8 ± 2      & 34 ± 9     &   6.96      \\ % 0.02 ± 0.02
                                 & mosaic      & 2,974     & 9 ± 2      & 48 ± 11    &   9.83      \\ % 0.02 ± 0.02
    \bottomrule

    \multirow{2}{*}{TCGA}         & collage     & 11,663    & 15 ± 3     & 74 ± 14    &   88.66       \\% 0.06 ± 0.07
                                 & mosaic      & 25,679    & 36 ± 20    & 157 ± 83   &   233.64    \\ % 0.16 ± 0.22 
    \bottomrule
    \end{tabular}
\end{table}

Table~\ref{tab:search_time} illustrates the significant storage and time factors that are contingent on the number of patches used in generating \emph{collages} and \emph{mosaics} within the context of image search. For Mayo-CRC and TCGA datasets, SPLICE consistently demonstrates its efficiency by utilizing less than half of the patches needed by Yottixel's mosaic. On average, only 10±2.43 and 15±2.78 patches per WSI were required for \emph{collage} generation in Mayo-CRC and TCGA, respectively, compared to 22±14.89 and 36±20.39 patches per WSI on average for \emph{mosaic} generation in Mayo-CRC and TCGA, respectively. The lower standard deviation for \emph{collage} generation highlights the reliability of SPLICE's approach by limiting the inclusion of redundant patches. In terms of storage requirements, using SPLICE for \emph{collage}-based image search results in a significantly reduced storage utilization. On average, the storage required for \emph{collage}-based search is 43±9.71 KB and 74±14.28 KB for Mayo-CRC and TCGA embeddings, respectively, as opposed to 96±51.29 KB and 157±83.29 KB for \emph{mosaic}-based search in Mayo-CRC and TCGA embeddings, respectively. These storage disparities reflect the efficiency of SPLICE in minimizing the data needed for effective image retrieval. The implications of these efficiency gains extend to search times. Yottixel's mosaic, due to its reliance on a larger pool of patches, necessitates more time for image retrieval. For instance, when retrieving similar images to 209 images in Mayo-CRC, Yottixel's mosaic requires a total search time of 16.89 seconds, compared to the 6.01 seconds needed by SPLICE. In the case of the TCGA dataset, using \emph{mosaic} for search consumes 3.9 minutes, while employing \emph{collage} reduces the retrieval time to just 1.5 minutes. Table~\ref{tab:search_time} also provides insights into the storage and time requirements for the Mayo-Liver dataset. Here, the difference between SPLICE and Yottixel is less pronounced, primarily because the available tissue area in the WSIs from Mayo-Liver is limited, making it not feasible to extract a large number of patches even when looser thresholds are applied. Nevertheless, SPLICE continues to exhibit its efficiency in balancing performance and resource consumption.

In summary, the experiments in this section underscore SPLICE's ability to efficiently represent histopathology images with a reduced number of patches, resulting in significant reductions in storage and search times, particularly in the context of large datasets. This efficiency positions SPLICE as a valuable tool in image search and retrieval, capable of delivering high-performance results with minimal resource utilization.

%----------------------------------------------------------------------
\section{Discussion}
%----------------------------------------------------------------------

The results of the experiments conducted to evaluate the performance of SPLICE in image search and retrieval, compared to the state-of-the-art method, namely Yottixel's mosaic~\cite{kalra2020yottixel}, have provided valuable insights into the efficacy and efficiency of SPLICE. The primary objective of these experiments was to compare the performance of SPLICE and Yottixel's mosaic in their ability to retrieve similar WSIs based on a query image. The critical differences between the two methods lie in their approach to patch selection. Yottixel's mosaic focuses on well-distributed patch selection, while SPLICE prioritizes the quality and uniqueness of selected patches. These differences have a significant impact on the results.

SPLICE's emphasis on selecting representative and less redundant patches was a key factor in its effective performance. This was evident in the experiments, where SPLICE consistently achieved better or comparable performance as Yottixel's mosaic with significantly fewer patches. Furthermore, one of the standout advantages of SPLICE is its efficiency. It requires significantly fewer patches for \emph{collage} generation compared to Yottixel's \emph{mosaic} approach. This not only reduces storage requirements but also leads to faster search times. The efficiency gains are particularly pronounced in larger datasets, making SPLICE an attractive option for applications where resource efficiency is crucial.

SPLICE's utilization of percentiles as thresholds for \emph{collage} generation offers a notable advantage in terms of flexibility when it comes to balancing performance and resource consumption. This approach requires only one parameter to be determined, the percentile, while Yottixel's mosaic, as the state-of-the-art method, necessitates adjusting two distinct parameters: the $k$ values for both color and spatial $k$-means clustering. This design choice simplifies the implementation and tuning process for its ease of use and reproducibility. Moreover, the impact of the SPLICE parameter on the performance is manageable, as the experiments demonstrated that SPLICE achieves satisfactory performance even with strict thresholds, allowing for a minimal number of patches. This flexibility can be valuable in various applications.

In conclusion, SPLICE's efficient and effective approach to image representation for search and retrieval has promising implications for various applications where histopathology images need to be processed. The experiments conducted in this study have demonstrated that SPLICE is a powerful representation for image search and retrieval. Its focus on patch quality, efficiency in resource utilization, and fewer parameters to adjust makes it a valuable tool for a wide range of applications. SPLICE's consistent and robust performance across diverse datasets underscores its potential to enhance histopathology image representation with a minimal representative set of patches.

%----------------------------------------------------------------------
\section{Methods}
%----------------------------------------------------------------------

\noindent\textbf{SPLICE: Color Characterization:} In the initial stage of SPLICE, the histopathology image is divided into a number of small patches. This step allows for the efficient handling of the image, breaking it down into more manageable, localized components that can be processed by available computational devices. Only patches containing tissue are included in subsequent processing steps (image background excluded). This is done by gathering patches from all segmented tissue areas of a given histopathology image (refer to Algorithm~\ref{alg:splice}, Lines \ref{line:segmentation_start}-\ref{line:segmentation_end}). This selective inclusion ensures that the analysis focuses on tissue regions while disregarding areas with minimal or no tissue content.

Within this selected subset of patches (that contain tissue), the color distribution within the RGB channels (red, green, and blue) of each individual patch is quantified (see Figure \ref{fig:collage}). Staining is of paramount importance, as it often serves as an indicator of variations in tissue composition. The analysis of color features allows us to identify patches that may exhibit monotonous or redundant color characteristics. Color features can be computed at a lower resolution, such as a 1.25x magnification, as higher resolutions do not necessarily provide a significant advantage for some tasks involving histopathology whole slide images~\cite{coudray2018classification,zarella2018estimation,tougaccar2020breastnet}. However, we can change to higher magnification after patch selection.   

\noindent\textbf{SPLICE: Sequential Analysis}: The second stage of SPLICE introduces a sequential analysis of color features, aiming to filter out patches with repetitive color characteristics (refer to Algorithm~\ref{alg:splice}, Lines \ref{line:collage_start}-\ref{line:collage_end}). As described in Figure ~\ref{fig:sequential}, this iterative process assesses the similarity between the color features of patches across a series of sequential scans. The color feature vector from the previous stage along with the standard deviation of every color channel is used for each scan: a single patch assumes the role of a reference point when a new scan cycle is initiated, against which the Euclidean distance to every other still non-excluded patch is calculated. 

The essence of this sequential color feature analysis lies in its capacity to identify and subsequently exclude patches that exhibit the color characteristics of a reference patch. This step is crucial because regions featuring redundant color attributes typically contribute limited additional information to the analysis. By systematically comparing patches through sequential scans, SPLICE ensures the systematic elimination of patches that do not significantly contribute to the overall representation of the histopathology image.

\begin{algorithm}
\caption{SPLICE for unsupervised Divide of a WSI to generate ``collage''}
\label{alg:splice}
\begin{algorithmic}[1]
\Function{SPLICE}{$\text{image}$}
    \State $\text{tissue\_area} \gets \text{segment\_tissue\_area}(\text{image})$ \label{line:segmentation_start}
    \State $\text{patches} \gets \text{patch\_tissue\_area}(\text{tissue\_area})$ \label{line:segmentation_end}
    
    \State $\text{collage} \gets \emptyset$ \label{line:collage_start}
    \State $\text{excluded\_patches} \gets \emptyset$
    \State $\text{percentile\_thresh} \gets k$
    \For{$\text{patch\_$i$ in patches}$}
        \If{$\text{patch\_$i$ not in excluded\_patches}$}
            \State $\text{reference\_patch} \gets \text{patch\_$i$}$
            \State $\text{color\_features} \gets \text{calculate\_color\_features}(\text{reference\_patch})$
            
            \For{$\text{patch\_$j$ in patches}$}
                \If{$\text{patch\_$j$ not in excluded\_patches}$}
                    \State $\text{distance} \gets \text{euclidean\_dist}(\text{color\_features, patch\_$j$.color\_features})$
                    \State $\text{threshold} \gets \text{$k^{th}$\_percentile\_distance}(\text{collage})$
                    
                    \If{$\text{distance} < \text{threshold}$}
                        \State $\text{excluded\_patches.append(patch\_$j$)}$
                    \EndIf
                \EndIf
            \EndFor
            \State $\text{collage.append(reference\_patch)}$
        \EndIf
    \EndFor \label{line:collage_end}
    
    \State \textbf{return} $\text{collage}$
\EndFunction
\end{algorithmic}
\end{algorithm}

In each scan, the $k^{th}$ percentile of the Euclidean distances between the reference patch and the remaining patches is computed. A Small $k$ value results in a more lenient similarity threshold and thus more patches included in the \emph{collage}. In contrast, a large $k$ value means a more strict similarity threshold to the reference patch, and therefore fewer selected patches. The $k^{th}$ percentile serves as a dynamic threshold, governing the exclusion of patches with color features closely mirroring those of the reference patch. The dynamic nature of this threshold adapts to the specific color distribution of each reference patch, facilitating nuanced and adaptive patch selection. Patches with a distance smaller than the percentile are systematically excluded. This iterative process continues until no further patches can be compared, resulting in the generation of the final SPLICE \emph{collage}, a representative view of the histopathology image. A standard \emph{collage} obtained after sequential analysis can be considerably smaller than the specimen area depicted in a given WSI. 

\noindent\textbf{Collage Indexing: Deep Feature Extraction}: After SPLICE has divided WSIs into a small set of representative and less redundant patches, namely the collage, the indexing process can be applied to each individual patch within the generated collage utilizing a pre-trained deep neural network. This operation should yield distinctive deep features, effectively encapsulating the intricate high-level tissue characteristics inherent in the collage. Of course, this will directly depend on the quality of embeddings acquired from a pre-trained or fine-tuned network. These collage representations emerge as pivotal elements for an array of downstream analytical tasks, including the facilitation of image search and classification. The utilization of the SPLICE collage for comparing and matching of histopathology images enables expeditious and resource-efficient image comparisons, leading to substantial reductions in search time and storage requirements due to the compact nature of the collage. This can be further magnified when \emph{binary} representations, like barcodes proposed by the Yottixel search engine \cite{kalra2020yottixel}, are used. The binarization method known as \emph{MinMax}~\cite{tizhoosh2015barcode,kumar2018deep} can be employed. This method converts the rich collage features into more efficient binary codes, enabling rapid real-time searches using the Hamming distance metric. This approach replaces the need for exhaustive k-nearest neighbor searches employing distance metrics like {$l_2$}, significantly optimizing search performance, particularly if implemented using logarithmic data types. For an average WSI file of approximately 700 MB, the binary codes can be compressed to as small as around 10 KB, resulting in a remarkable reduction in size, approximately 70,000 times smaller than the original file~\cite{kalra2020yottixel}. 

With a sufficiently extensive archive of indexed WSI collages, SPLICE can emerge as a powerful and interactive interface, facilitating the seamless generation and indexing of query WSIs while adeptly matching them with the most similar WSIs in the archive, all guided by the collage representations. This transformative process not only streamlines research and analysis but also empowers users to navigate and extract insights from vast repositories of WSIs with unprecedented efficiency and efficacy.

\bibliography{sn-bibliography}

\end{document}